\documentclass[%
 aip,graphicx]{revtex4-2}

\usepackage{graphicx}
\usepackage{bm}
\usepackage[version=3]{mhchem} 
\usepackage{xcolor} 

\usepackage[normalem]{ulem}

\usepackage{times}

\usepackage[exponent-product = \cdot, separate-uncertainty = true, multi-part-units=single]{siunitx}
\usepackage{graphicx}
\usepackage{amsmath}

\begin{document}

\preprint{APS/123-QED}

\title{Few-femtosecond electronic and structural rearrangements of CH$_4^+$ driven by the Jahn-Teller effect}

\author{Kristina S. Zinchenko}
\thanks{These authors contributed equally to this work}
\affiliation{Laboratory of Physical Chemistry, ETH Z\"{u}rich, 8093 Z\"{u}rich, Switzerland}
\author{Fernando Ardana-Lamas}
\thanks{These authors contributed equally to this work}
\affiliation{Laboratory of Physical Chemistry, ETH Z\"{u}rich, 8093 Z\"{u}rich, Switzerland}
\author{Valentina Utrio Lanfaloni}
\thanks{These authors contributed equally to this work}
\affiliation{Laboratory of Physical Chemistry, ETH Z\"{u}rich, 8093 Z\"{u}rich, Switzerland}
\author{Nicholas Monahan}
\affiliation{Laboratory of Physical Chemistry, ETH Z\"{u}rich, 8093 Z\"{u}rich, Switzerland}
\author{Issaka Seidu}
\affiliation{National Research Council of Canada, Ottawa, ON, Canada}
\author{Michael S. Schuurman}
\affiliation{National Research Council of Canada, Ottawa, ON, Canada}
\author{Simon P. Neville}
\email{simon.neville@nrc-cnrc.gc.ca}
\affiliation{National Research Council of Canada, Ottawa, ON, Canada}
\author{Hans Jakob Wörner}
\email{hwoerner@ethz.ch}
\affiliation{Laboratory of Physical Chemistry, ETH Z\"{u}rich, 8093 Z\"{u}rich, Switzerland}

\date{\today}

\begin{abstract}
The Jahn-Teller effect (JTE) is central to the understanding of the physical and chemical properties of a broad variety of molecules and materials. Whereas the manifestations of the JTE on stationary properties of matter are relatively well studied, the study of JTE-induced dynamics is still in its infancy, largely owing to its ultrafast and non-adiabatic nature. For example, the time scales reported for the distortion of CH$_4^+$ from the initial $T_{\rm d}$ geometry to a nominal $C_{\rm 2v}$ relaxed structure range from 1.85~fs over 10$\pm$2~fs to 20$\pm$7~fs. Here, by combining element-specific attosecond transient-absorption spectroscopy and quantum-dynamics simulations, we show that the initial electronic relaxation occurs within 5~fs and that the subsequent nuclear dynamics are dominated by the $Q_2$ scissoring and $Q_1$ symmetric stretching modes, which dephase in 41$\pm$10~fs and 13$\pm$3~fs, respectively. Significant structural relaxation is found to take place only along the e-symmetry $Q_2$ mode. These results demonstrate that CH$_4^+$ created by ionization of CH$_4$ is best thought of as a highly fluxional species that possesses a long-time-averaged vibrational distribution centered around a $D_{\rm 2d}$ structure. The methods demonstrated in our work provide guidelines for the understanding of Jahn-Teller driven non-adiabatic dynamics in other, more complex systems.
\end{abstract}

\maketitle


\section{Introduction}

The Jahn-Teller effect (JT, JTE) plays a fundamental role in the understanding of the structure and dynamics of molecules, metal complexes, and solids. In its original formulation, the JT theorem states that "a nonlinear polyatomic system in a spatially degenerate electronic state distorts spontaneously in such a way that the degeneracy is lifted and a new equilibrium structure of lower symmetry is attained." \cite{jahn1937stability}.
The JTE is indeed responsible for the distortion of the geometric structure of open-shell molecules\cite{woerner09b}, such as charged fullerenes \cite{gunnarsson95a,chancey97a,dunn2005jahn}, metal complexes\cite{bersuker20a,streltsov20a} and perovskites\cite{varignon19a}, but it also plays a role in the explanation of superconductivity and colossal magnetoresistance \cite{ramirez97a,halcrow13a}. The JTE is a consequence of a strong coupling between electronic and nuclear dynamics, also known as vibronic coupling, and the highest-symmetry configuration of a JT-active molecule corresponds to a conical intersection \cite{domcke04a}. The topology of potential-energy surfaces (PES) of JT-active systems, therefore, induce ultrafast dynamics that are representative of systems featuring conical intersections \cite{schuurman18a}.

The ionization of highly symmetric, closed-shell molecules offers the interesting opportunity of preparing a molecular wave packet centered at the location of a conical-intersection seam and observing the ultrafast coupled electronic and nuclear dynamics initiated by suddenly turning on the Jahn-Teller effect. The methane cation (CH$_4^+$) offers a particularly interesting example of such dynamics because its electronic ground state is triply degenerate at the $T_{\rm d}$ geometry of CH$_4$, such that ionization of CH$_4$ prepares the cation at the location of a three-fold conical intersection seam. 

Here, we use element-specific attosecond transient-absorption spectroscopy (ATAS) at the carbon K-edge\cite{Zinchenko:20} to observe the structural and non-adiabatic dynamics of CH$_4^+$ driven by the Jahn-Teller effect. The measurements are interpreted by comparison with quantum-dynamical simulations in full dimensionality. Such a description is necessary to obtain a qualitatively correct description of the dynamics because of the high degree of correlation between the vibrational degrees of freedom\cite{frey88a}, combined with the geometric-phase effects \cite{woerner06c} that cannot be neglected. Previous work has shown that the PES of CH$_4^+$ possesses 12 equivalent minima of $C_{\rm 2v}$ symmetry, 8 two-fold conical intersections of $C_{\rm 3v}$ symmetry between the two lowest adiabatic sheets of the PES and 6 saddle-points of $D_{\rm 2d}$ symmetry on the lowest adiabatic PES\cite{frey88a}. The presence of a geometric phase fundamentally modifies the dynamics of CH$_4^+$. This is known from high-resolution photoelectron spectroscopy, which has established that the sequence of the lowest-lying vibronic levels is $t_2$ below $t_1$, whereas a treatment that ignores the geometric phase predicts the qualitatively different energetic sequence of vibronic levels $a_1$, $t_2$, $e$ \cite{woerner06c,woerner07b}. This indicates that the geometric phase can be expected to profoundly modify the structural rearrangement of CH$_4^+$ following ionization.

Previous experimental works that have addressed the JT dynamics in the methane cation include high-harmonic spectroscopy \cite{Baker2006}, time-resolved strong-field ionization (SFI) \cite{Li2021} and transient-absorption spectroscopy\cite{ridente23a}. High-harmonic spectroscopy has been used to obtain the ratio of the nuclear auto-correlation functions of CD$_4$ and CH$_4$ over the first 1.7~fs following SFI. This has motivated theoretical work\cite{mondal11a,mondal14a}, which has eventually concluded that it takes CH$_4^+$ only 1.85~fs following SFI to attain a nominal $C_{\rm 2v}$ structure\cite{mondal15a}. Time-resolved SFI with 25~fs pulses, combined with a two-dimensional description of the dynamics concluded that it takes 20$\pm$7~fs for CH$_4^+$ to reach its $C_{\rm 2v}$ equilibrium geometry. A recent quantum-dynamical study performed on two-dimensional PES of CH$_4^+$ concluded that CH$_4^+$ first adopted a $D_{\rm 2d}$ structure before reaching the $C_{\rm 2v}$-symmetric minimum \cite{goncalves21a,blavier21a}. Most recently, ATAS combined with classical-trajectory calculations concluded that it takes CH$_4^+$ 10$\pm$2~fs to reach its $C_{\rm 2v}$ equilibrium geometry\cite{ridente23a}. 

In the present work, we show that SFI of methane prepares CH$_4^+$ at a three-fold conical intersection, from where the population is found to relax to the lowest adiabatic surface within 5~fs (3.9$\pm$0.4~fs from a monoexponential fit), inducing large-amplitude, multi-mode vibrational dynamics. The initial nuclear dynamics are dominated by the $Q_1$ symmetric stretch of $a_1$ and the $Q_2$ scissoring mode of $e$ symmetry, which distorts CH$_4^+$ from its initial $T_d$ geometry to a $D_{\rm 2d}$ geometry. Significant structural relaxation is found to take place only along one coordinate of this $e$-symmetry mode. We moreover find a pronounced multimode character of the vibrational dynamics by displaying the characteristic frequencies of both the stretching and the scissoring modes, which are damped on different time scales.

\section{Methods}
\subsection{Experimental setup}
\label{experiment}
The experimental setup consists of a cryogenically cooled 1 kHz Ti:Sa laser (Coherent) that pumps an optical parametric amplifier (Light Conversion) to produce passively CEP-stable 2.5 mJ pulses centered at 1.76 \textmu m. They are broadened in an argon-filled hollow-core fiber \cite{silva15a} and post-compressed with bulk material down to sub-two optical cycles (temporal pulse duration of 10.4 $\pm$ 1.5 fs) \cite{Zinchenko:20, Zinchenko:22}.  A beamsplitter then splits the hollow-core-fiber output into two arms: the transmitted beam is used as a pump to excite the sample by SFI; the reflected one is focused on a helium-filled finite gas cell, where the high-harmonic-generation process occurs and produces an isolated attosecond ($<$200~as) soft X-ray (SXR) pulse with a cutoff energy of $\sim$400~eV. The SXR beam is then focused by a toroidal mirror into the sample and the transmitted photons detected by a CCD-camera-based spectrometer are used as the system's probe. To look at the temporal evolution after ionization, the pump, and probe are delayed with respect to each other by a delay stage integrated into the optical pump beam path. More details on the experimental setup are given in \cite{Zinchenko:22, Zinchenko2023}.

\subsection{Data analysis}
\subsubsection{Static spectra and change in optical density}

The optical density (OD), shown in Figure \ref{fig:1}\textbf{b} (red line) is defined as:
\begin{center}
    $OD(\hbar\omega)=log_{10}\frac{I_0(\hbar\omega)}{I_{pump ~off}(\hbar\omega)}$
\end{center}

where $I_0(\hbar\omega)$ is the reference spectral intensity, i.e., the spectrum recorded without sample, and $I_{pump~off}$ is the spectrum of the non-ionized sample. Both spectra ($I_0$ and $I_{pump~off}$ are background-corrected.\\
The spectral calibration of the image, acquired with our spectrometer, is based on the absorption bands of ethylene at the carbon K-edge (287.4, 288.66 eV and 284.3, 285.15 \cite{HITCHCOCK19801} and 288.5 eV due to the carbon contamination of the SXR reflective optics) and the nitrogen K-edge of N$_4$ (at 400.0 eV \cite{HITCHCOCK19801}) as a reference at higher photon energy.\\

The change in the optical density given by the pump is calculated at each time delay $\tau$ as follows:
\begin{center}
     $\Delta OD(\tau)=-  \log_{10} \frac{I_{pump~on}(\tau)}{I_{pump~off}}$
\end{center}
where $I_{pump~on}(\tau)$ and $I_{pump~off}$ are the spectra collected with and without the delayed MIR beam, respectively.

\subsubsection{Vibrational analysis}
For the vibrational analysis (see Fig. \ref{fig:2}c), the absorption band centered at \SI{281}{\eV} has been isolated from the experimental and theoretical $\Delta$OD($\hbar\omega,\tau_d$) datasets. For every spectrum at each time step the center of mass of the isolated absorption bands has been calculated. The obtained center of mass as a function of time delay has been Fourier-transformed with a Blackman-Harris window and zero-padding four times the length of the datasets.

The Gabor transform analysis (see Figs.\ref{fig:4} and \ref{fig:4b} has been done with the following method. First, the center of mass of the time-dependent nuclear density for each vibrational mode ($Q_1$, $Q_{2x}$, $Q_{2y}$, $Q_{3x}$, $Q_{3y}$, $Q_{3z}$, $Q_{4x}$, $Q_{4y}$, $Q_{4z}$) has been calculated at each time step. Then, the experimental and theoretical calculated center of mass from the $\Delta$OD($\hbar\omega,\tau_d$) datasets and the ones from the time-dependent nuclear densities have been Gabor transformed. As a special case of short-time Fourier transforms, the signals have been divided into shorter segments of equal length, multiplied by a Gaussian function, and the resulting function has been Fourier transformed with zero-padding to derive the time-frequencies analysis. To visualize the change of the nonstationary signal's frequencies over time, the spectrograms of each Gabor transforms have been reported.

\subsection{First-principles calculations}
\label{theory}
\subsubsection{Model Hamiltonian}
The total molecular Hamiltonian, $\hat{H}$, was represented in a basis $\{|I\rangle\}$ of quasi-diabatic electronic states:

\begin{equation}
    \begin{aligned}
        \hat{H} &= \sum_{I,J} | I \rangle \langle I | \hat{H} | J \rangle \langle J | \\
        &= \sum_{I} | I \rangle T_{II} \langle I | + \sum_{I,J} | I \rangle W_{IJ}(\boldsymbol{Q}) \langle J |,
    \end{aligned}
\end{equation}

\noindent
Here, the nuclear kinetic energy operator matrix $\boldsymbol{T}$ in terms of dimensionless mass- and frequency-scaled normal modes $Q_{\alpha}$ as

\begin{equation}
    \boldsymbol{T} = \left( -\frac{1}{2} \sum_{\alpha} \omega_{\alpha} \frac{\partial^{2}}{\partial Q_{\alpha}^{2}} \right) \boldsymbol{1} = \hat{T} \boldsymbol{1},
\end{equation}

\noindent
where $\omega_{\alpha}$ is the frequency for mode $Q_{\alpha}$. The nuclear-coordinate-dependent quasi-diabatic potential matrix $\boldsymbol{W}(\boldsymbol{Q})$ has elements

\begin{equation}
    W_{IJ}(\boldsymbol{Q}) = \langle I | \hat{H}_{el} | J \rangle,
\end{equation}

\noindent
where $\hat{H}_{el}$ denotes the electronic Hamiltonian; $\hat{H}_{el} = \hat{H} - \hat{T}$. The electronic states $|I\rangle$ pertinent to the dynamics of CH$_{4}$ following SFI, and the consequent probing of these via X-ray absorption, are: (i) the those spanning the triply-degenerate cationic ground state manifold, which we denote by $\{ | \tilde{X}_{i}^{+} \rangle | i \in \{x,y,z\} \}$, and; (ii) the singly-degenerate first core-ionised state, denoted by $| \tilde{\mathcal{C}}^{+} \rangle$, corresponding to the $1s \rightarrow \text{HOMO}$ transition. For brevity, let the potential matrix elements be abbreviated as follows:

\begin{equation}
    \langle \tilde{X}_{i}^{+} | \hat{H}_{el} | \tilde{X}_{j}^{+} \rangle = W_{ij}, \hspace{0.5cm} i,j \in \{x,y,z\},
\end{equation}

\begin{equation}
    \langle \tilde{\mathcal{C}}^{+} | \hat{H}_{el} | \tilde{\mathcal{C}}^{+} \rangle = W_{cc}.
\end{equation}

\noindent
Then, the matrix representation of the field-free Hamiltonian reads

\begin{equation}
\boldsymbol{H} = \hat{T} \boldsymbol{1} +
    \begin{bmatrix}
        W_{xx} & W_{xy} & W_{xz} & 0 \\
        W_{yx} & W_{yy} & W_{yz} & 0 \\
        W_{zx} & W_{zy} & W_{zz} & 0 \\
        0      & 0      & 0      & W_{cc}
    \end{bmatrix}.
\end{equation}

Each element of the nuclear-coordinate-dependent quasi-diabatic potential matrix $\boldsymbol{W}(\boldsymbol{Q})$ must be cast into an (approximate) closed analytical form for use in quantum dynamics simulations. For this, we use the vibronic coupling Hamiltonian model of K\"{o}ppel, Domcke and Cederbaum\cite{cederbaum1981,koppel1984}, in which each potential matrix element is Taylor expanded in terms of the ground state normal modes $Q_{\alpha}$ about the ground state minimum energy geometry $\boldsymbol{Q}_{0}$. In our model, we expand each matrix element to 4th-order in with respect to the one-mode terms and to 2nd-order with respect to the two-mode terms:

\begin{equation}\label{eq:modpot}
    W_{IJ}(\boldsymbol{Q}) \approx W_{IJ}(\boldsymbol{Q}_{0}) + \sum_{\alpha} \sum_{n=1}^{4} \frac{1}{n!} \tau_{\alpha,n}^{(I,J)} Q_{\alpha} + \frac{1}{2} \sum_{\alpha,\beta} \eta_{\alpha \beta}^{(I,J)} Q_{\alpha} Q_{\beta}.
\end{equation}

The global gauge of the adiabatic-to-diabatic transformation was fixed by taking the two representations to be equal at the point of expansion, $\boldsymbol{Q}_{0}$, yielding

\begin{equation}
    W_{IJ}(\boldsymbol{Q}_{0}) = \delta_{IJ} V_{I}(\boldsymbol{Q}_{0}),
\end{equation}

\noindent
where $\{ V_{I} \}$ denotes the set of adiabatic potential energies. For the valence-ionized block of $\boldsymbol{W}$, the remaining expansion coefficients $\{ \tau_{\alpha,n}^{(I,J)}, \eta_{\alpha\beta}^{(I,J)} \}$ were determined via direct least squares fitting to quasi-diabatic potential matrix element values computed using a propagative variant of the block diagonalization diabatisation (P-BDD) method\cite{neville2020}. See Appendix~C of Reference~\citenum{neville2020} for a full description of the fitting procedure. The P-BDD procedure requires as input adiabatic energies and electronic wave function overlaps. These were computed at the multi-reference configuration interaction (MRCI) level of theory using the cc-pVTZ basis set. The reference space used corresponds to a complete active space (CAS) formed from the $2s$ and $2p$ orbitals. The orbital basis was optimized at the CAS self-consistent field (CASSCF) level of theory using this active space in conjunction with state averaging. The final MRCI wave functions were constructed by allowing all single excitations out of the CAS reference space. As the core-ionized state $| \tilde{\mathcal{C}}^{+} \rangle$ is energetically well-separated from its orthogonal complement, the quasi-diabatic potential matrix element $W_{cc}(\boldsymbol{Q})$ may be equated with the corresponding adiabatic potential energy surface. The expansion coefficients for this matrix element were computed via fitting to adiabatic energies computed at the ionization potential equation of motion coupled cluster singles and doubles (EOM-IP-CCSD) level of theory within the core-valence separation approximation (CVS-EOM-IP-CCSD)\cite{vidal2019} using the cc-pVTZ basis. The MRCI calculations were performed using the COLUMBUS set of programs\cite{columbus}, and the CVS-EOM-IP-CCSD calculations using the QChem program\cite{qchem}.

\subsubsection{Wave packet propagations}
Full (9-dimensional) wave packet propagations simulating the non-adiabatic dynamics following SFI to the $D_{0}$ manifold were performed using the multi-configurational time-dependent Hartree (MCTDH) method\cite{meyer90a,manthe1992,beck2001,meyer09a}. The so-called multi-set formalism was used, in which (using a useful abuse of notation) the wave packet \textit{ansatz} reads

\begin{equation}
    | \Psi(t) \rangle = \sum_{I} | I \rangle | \Psi_{I}(\boldsymbol{Q}, t) \rangle,
\end{equation}

\begin{equation}
    | \Psi_{I}(\boldsymbol{q}, t) \rangle = \sum_{j_{1}=1}^{n_{1}^{(I)}} \cdots \sum_{j_{f}=1}^{n_{f}^{(I)}} A_{j_{1},\dots,j_{f}}^{(I)} \bigotimes_{\kappa=1}^{f} | \varphi_{j_{\kappa}}^{(\kappa;I)}(q_{\kappa},t) \rangle.
\end{equation}

\noindent
Here, the so-called single-particle functions (SPFs) $\varphi_{j}^{(\kappa;I)}(q_{\kappa},t)$ are each functions of logical/combined modes $q_{\kappa}$, each being a generally multidimensional subset of $d_{\kappa}$ physical coordinates $Q_{\alpha}$: $q_{\kappa} = (Q_{i_{1}^{\kappa}, \dots, Q_{i_{d_{\kappa}^{\kappa}}}})$. The SPFs are further expanded in terms of a primitive discrete variable representation (DVR) basis, chosen here as a harmonic oscillator DVR. The mode combination scheme used as well as the numbers of SPF and DVR basis functions used are given in the Supplementary Information along with the normal modes vectors.

The initial wave packet $|\Psi(t=0)\rangle$ was taken to correspond to vertical ionization of the neutral ground state $|\tilde{X}\rangle$ to the valence-ionized manifold:

\begin{equation}
    | \Psi(t=0) \rangle = \sum_{i\in\{x,y,z\}} \left( | \tilde{X}_{i}^{+} \rangle \langle \tilde{X} | + h.c. \right) | \Psi_{GS} \rangle,
\end{equation}

\noindent
where $| \Psi_{GS} \rangle$ denotes the neutral ground vibronic eigenstate, constructed here within the harmonic approximation.

All MCTDH calculations were performed using the Quantics quantum dynamics code\cite{worth2020,quantics}.

\subsubsection{ATAS simulation}
The simulated ATAS at time $\tau$, denoted by $\sigma(\tau, \omega)$ here, was constructed by vertically exciting the time-evolving wave packet in the $D_{0}$ manifold to the core-ionized state $| \tilde{\mathcal{C}}^{+} \rangle$, continuing the propagation and then Fourier transforming the resulting wave packet autocorrelation function:

\begin{equation}\label{eq:atas_sim}
    \sigma(\tau, \omega) \sim \int_{\tau}^{\infty} \langle \Psi(\tau) | \hat{\epsilon}^{\dagger} e^{-i\hat{H}(t-\tau)} \hat{\epsilon} | \Psi(\tau) \rangle e^{i \omega (t-\tau)} dt,
\end{equation}

\noindent
where atomic units have been assumed and

\begin{equation}\label{eq:epsilon}
    \hat{\epsilon} = \sum_{i \in \{x,y,z\}} | \tilde{\mathcal{C}}^{+} \rangle \langle \tilde{X}_{i}^{+} |.
\end{equation}

This corresponds to a perturbative description of the ATAS spectrum assuming $\delta$-function laser pulses, and is analogous to the methodology developed by Richings and Worth for the simulation of time-resolved photoelectron spectra\cite{richings2014}. We note that the form of the operator $\hat{epsilon}$ in Equation~\ref{eq:epsilon} corresponds to the adoption of the Condon approximation in the description of the interaction with the probe pulse. This may be justified when used in conjunction with a diabatic basis, for which transition dipoles can be expected to have a relatively weak nuclear coordinate dependence.

To ameliorate artifacts arising from the use of a finite propagation time $T=200$~fs following projection onto the core-ionized state, the integrand in Equation~\ref{eq:atas_sim} was multiplied by the following window function:

\begin{equation}
    g(t) = \cos^{2} \left[ \frac{\pi (t-\tau)}{2T} \right] \Theta \left(1 - \frac{|t-\tau|}{T} \right),
\end{equation}

\noindent
where $\Theta$ denotes the Heaviside step function.

\section{Results}

 \begin{figure}[!htb]]
    \includegraphics[width=1.0\textwidth]{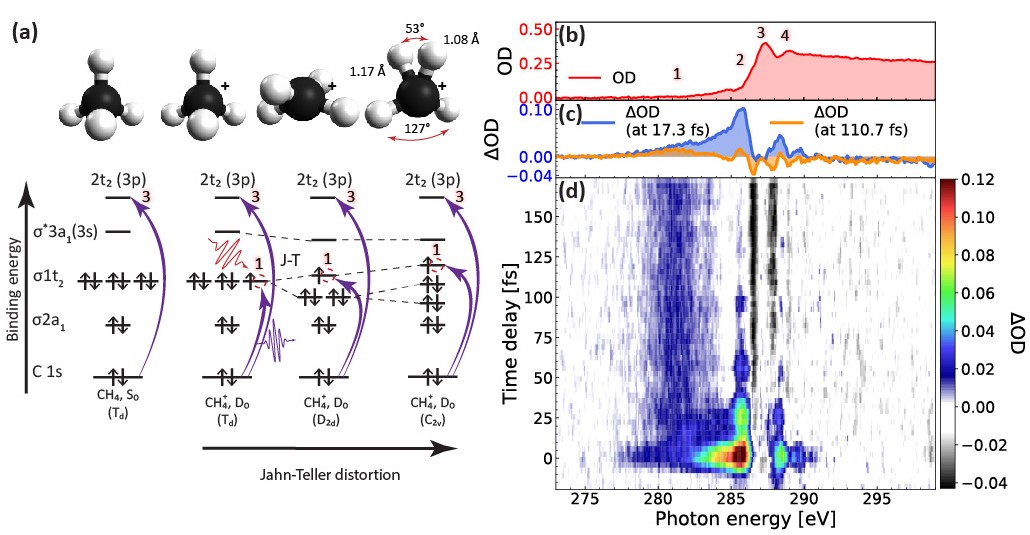}
    \caption{\textbf{Attosecond transient-absorption spectroscopy of methane cation at the carbon K-edge.} (\textbf{a}) Molecular-orbital diagram of methane and methane cation in their initial T$_{\rm d}$ geometry and CH$_4^+$ in its JT-distorted D$_{\rm 2d}$ and C$_{\rm 2v}$ geometries. (\textbf{b}) Static X-ray absorption spectrum of methane (\textbf{c}) Difference spectra ($\Delta$OD) between pumped and unpumped samples at early and late time delays (\textbf{d}) Difference spectra as a function of the pump-probe delay. 
    \label{fig:1}
     }
\end{figure}

A carrier-envelope-phase-(CEP)-stable sub-two-cycle (10.4$\pm$1.5~fs) laser pulse centered at 1.76~$\mu$m is employed to strong-field ionize CH$_4$, producing CH$_4^+$, the dynamics of which are recorded by attosecond transient-absorption spectroscopy in a dispersive geometry using an isolated attosecond pulse covering the carbon K-edge. Details on the experimental setup are given in the Section 
\ref{experiment}.

Figure \ref{fig:1} provides an overview of the experimental results and the assignment of the observed transitions. Panel \textbf{a} shows the relevant structures and molecular orbitals of CH$_4$ and CH$_4^+$. Panel \textbf{b} shows the experimental X-ray absorption spectrum of neutral methane in its electronic ground state. 
The strongest transition at \SI{288.0}{\eV} (labeled "3") corresponds to the transition C1s$\rightarrow$3p (2t$_2$). The following structure up to \SI{288.7}{\eV} ("4") is assigned to C1s$\rightarrow n$p~(t$_2$) with $n\geq$ 4 transitions. The absorption feature at \SI{288.7}{\eV} and the following structures are assigned to C1s$\rightarrow$3d transitions (split by the $T_{\rm d}$ geometry). A weak absorption feature centered at \SI{287.0}{\eV} ("2") is assigned to the C1s$\rightarrow$3s~(3a$_1$) Rydberg transition, which is detected because of vibronic coupling, i.e., this transition is accompanied by the excitation of vibrations of t$_2$ symmetry. The assignment of features 2-4 is based on previous work\cite{brown1978fine,urquhart2005rydberg}.

The changes in optical density induced by the pump pulse ($\Delta$OD) at two selected delays are shown in Fig.~\ref{fig:1}\textbf{c}. Both spectra have in common an additional absorption band ("1") centered around \SI{281}{\eV}, which is assigned to the C1s$\rightarrow$HOMO (highest-occupied molecular orbital) transition, i.e. C1s$\rightarrow$1t$_2$ at the $T_{\rm d}$ geometry. Figure \ref{fig:1}\textbf{d} shows $\Delta$OD as a function of the pump-probe delay, where a positive delay corresponds to the mid-infrared (MIR) pulse preceding the soft-X-ray (SXR) pulse.

\begin{figure}
    \includegraphics[width=1.0\textwidth]{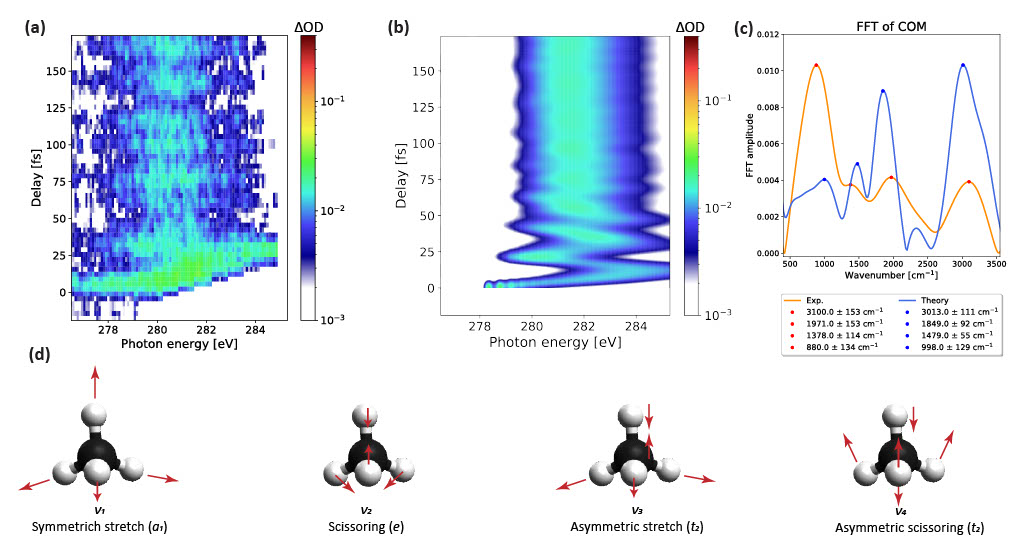}
\caption{\textbf{Structural encoding of CH$_4^+$ dynamics in ATAS.} (\textbf{a}) Measured $\Delta$OD as a function of pump-probe delay in the region of interest to the present work. (\textbf{b}) Calculated $\Delta$OD as a function of pump-probe delay. (\textbf{c}) Fourier transform of the center of mass of the absorption band centered at \SI{281}{\eV} for the measured data (orange) and theoretical calculations (blue). The assignments of these frequencies are discussed in the main text. (\textbf{d}) Vibrational modes of CH$_4^+$: $Q_1$ symmetric stretching (\textit{a$_1$}, 3029~cm$^{-1}$), $Q_2$ scissoring (\textit{e}, 1561~cm$^{-1}$), $Q_3$ asymmetric stretching (\textit{t$_2$}, 3133~cm$^{-1}$), $Q_4$ scissoring (\textit{t$_2$}, 1345~cm$^{-1}$). Only one component of each of the degenerate modes is shown.}
\label{fig:2}
\end{figure}

At long pump-probe delays the transient spectra are dominated by the C1s$\rightarrow$HOMO band centered at 281~eV that undergoes damped oscillations of its central position and intensity. At short delays, an additional absorption band is observed that extends from 283-287~eV, which rapidly decays into a narrow absorption band centered at 285.7~eV and displays damped periodic intensity oscillations. This part of the spectrum is assigned to dynamics induced by the strong MIR field in CH$_4$, similar to recent observations in SiH$_4$\cite{matselyukh22a}. Since our calculations were all performed on CH$_4^+$, not the neutral CH$_4$, these spectral features do not appear in the simulations.
For the remainder of this article, we will concentrate on the dynamics of CH$_4^+$, encoded in absorption band 1, which is reproduced in Fig.~\ref{fig:2}\textbf{a}.

These experimental results are interpreted through comparison with quantum-dynamics simulations of both the non-adiabatic dynamics following ionization to the $T_{\rm d}$ cationic ground state and the resulting ATAS spectra. These calculations were performed at the multi-configurational time-dependent Hartree (MCTDH) level of theory\cite{meyer90a,manthe1992,beck2001,meyer09a} using a vibronic-coupling Hamiltonian\cite{cederbaum1981,koppel1984} parameterized by fitting to \textit{ab-initio} quasi-diabatic potentials computed at the multi-reference configuration interaction (MRCI) and equation of motion coupled cluster singles and doubles (EOM-CCSD) levels of theory. Details of these calculations are given in Section \ref{theory}. In order to validate the ability of the model Hamiltonian to describe the complex non-adiabatic dynamics following ionization to the $D_{0}$ manifold, it was used to simulate the first band in the photoelectron spectrum of CH$_{4}$. The resulting spectrum is shown in the Supplementaty Material alongside the experimental spectrum of Potts and Price\cite{potts1972}. Overall, the two spectra are in excellent agreement providing some confidence of the ability of the model to correctly describe the dynamics of CH$_{4}$ following ionisation.

Figure~\ref{fig:2}\textbf{a} shows the measured $\Delta$OD in the region of \SIrange{275}{285}{eV}, that corresponds to the C1s$\rightarrow$HOMO transition. This feature undergoes large-amplitude periodic oscillations, in good agreement with the simulated $\Delta$OD reported in figure \ref{fig:2}\textbf{b}. This absorption band subsequently shifts from 278~eV to 284~eV in just 13~fs, shifts back to 280~eV by 20.5~fs, and then undergoes damped oscillations towards larger time delays.

A fast Fourier transform (FFT) of the center of mass of band 1 (Fig.~\ref{fig:2}\textbf{c}, orange curve for the measured data, blue curve for the calculations) reveals four dominant frequencies: 3100$\pm$153~cm$^{-1}$, 1971$\pm$153~cm$^{-1}$, 1378$\pm$114~cm$^{-1}$ and 880$\pm$134~cm$^{-1}$. These frequencies correspond very well to those obtained from the center of mass of the calculated $\Delta$OD. The highest frequency is consistent with the calculated harmonic frequencies of the stretching vibrations ($Q_1$ and $Q_3$, illustrated in Fig. \ref{fig:2}\textbf{d}) and the third-highest is consistent with that of the $Q_2$ scissoring mode of e symmetry, but the second-highest and lowest frequencies do not correspond to any calculated harmonic frequencies. As we show below, the harmonic frequencies have to be used with care in assigning the observed dynamics because the strong vibronic coupling can induce dynamics with additional periodicities that do not correspond to the harmonic frequencies.

\begin{figure}[t]
    \centering
    \includegraphics[width=0.8\textwidth]{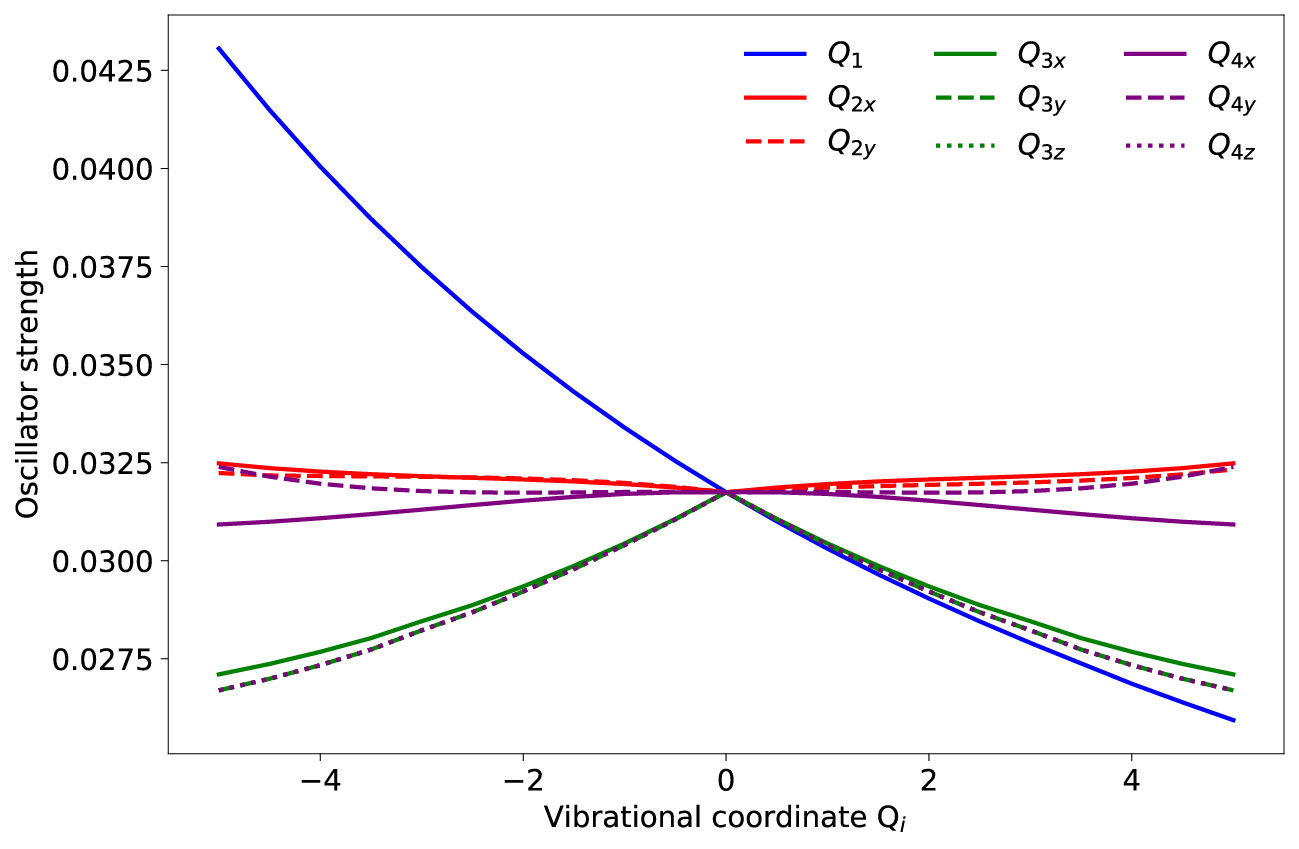}
    \caption{\textbf{Oscillator strengths} of the C1s$\rightarrow$HOMO transition along each of the 9 normal-mode coordinates of CH$_4^+$.
    \label{fig:3c}
    }
\end{figure}

\begin{figure}[!htbp]
    \includegraphics[width=0.8\textwidth]{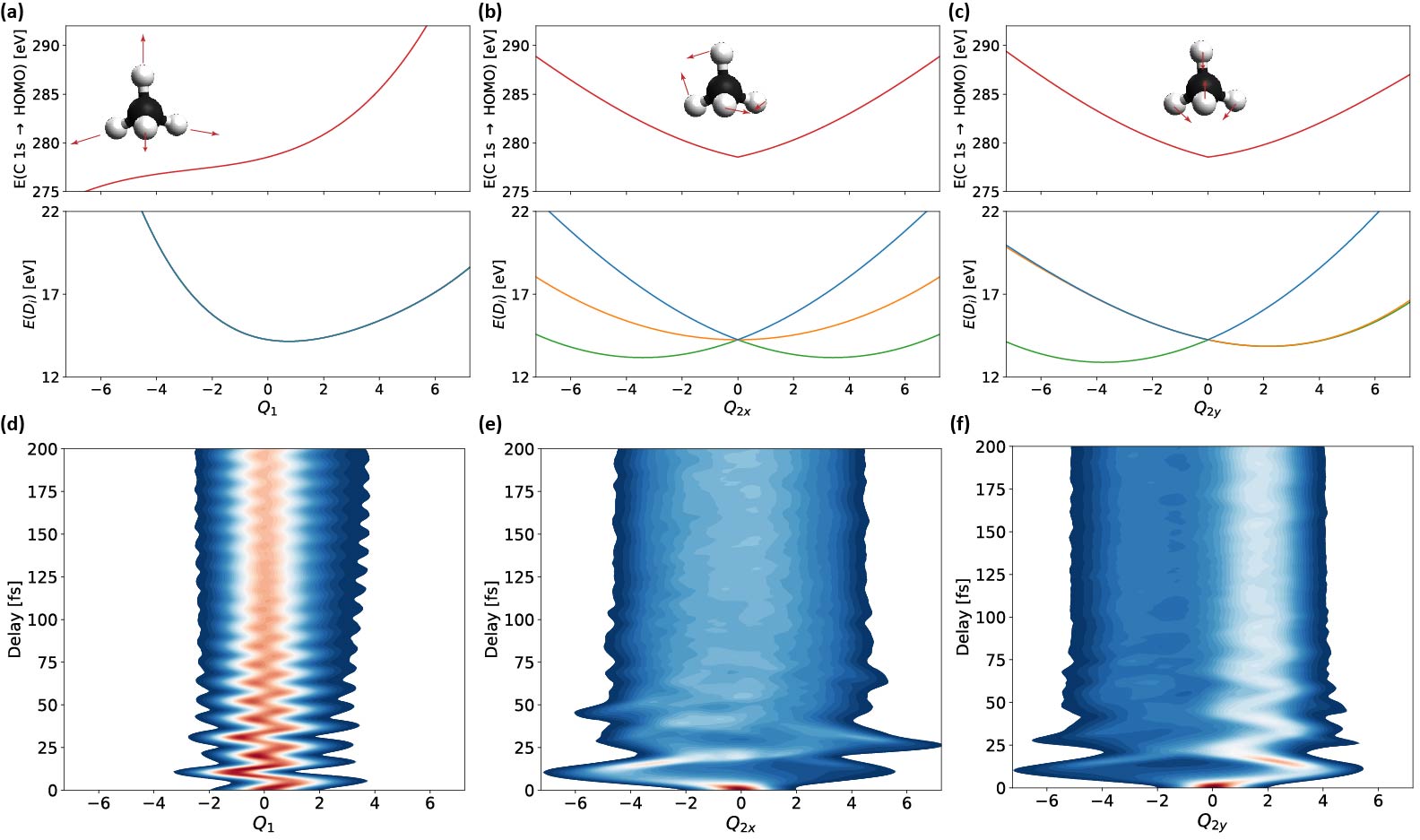}
    \caption{
    \textbf{Time evolution of the nuclear wave packet.}
    Geometry dependence of the core-excitation energies (top panels) and the three lowest-lying electronic states ($D_{0-2}$) of methane cation (bottom panels) for the $Q_1$ (symmetric stretch) (\textbf{a}) and Jahn-Teller active $Q_2$ mode (\textbf{b} and \textbf{c}). 
    The $Q_{2y}$ mode preserves $D_{\rm 2d}$ symmetry such that the lower (upper) potential-energy curves in \textbf{c}) are degenerate for $Q_{2y}>0$ ($Q_{2y}<0$).
    The corresponding time-dependent nuclear density for the $Q_1$, $Q_{2x}$, and $Q_{2y}$ modes are shown in panels \textbf{d}, \textbf{e}, and \textbf{f}, respectively.
    }
   \label{fig:3}
\end{figure}

\begin{figure}[t]
    \centering
    \includegraphics[width=\textwidth]{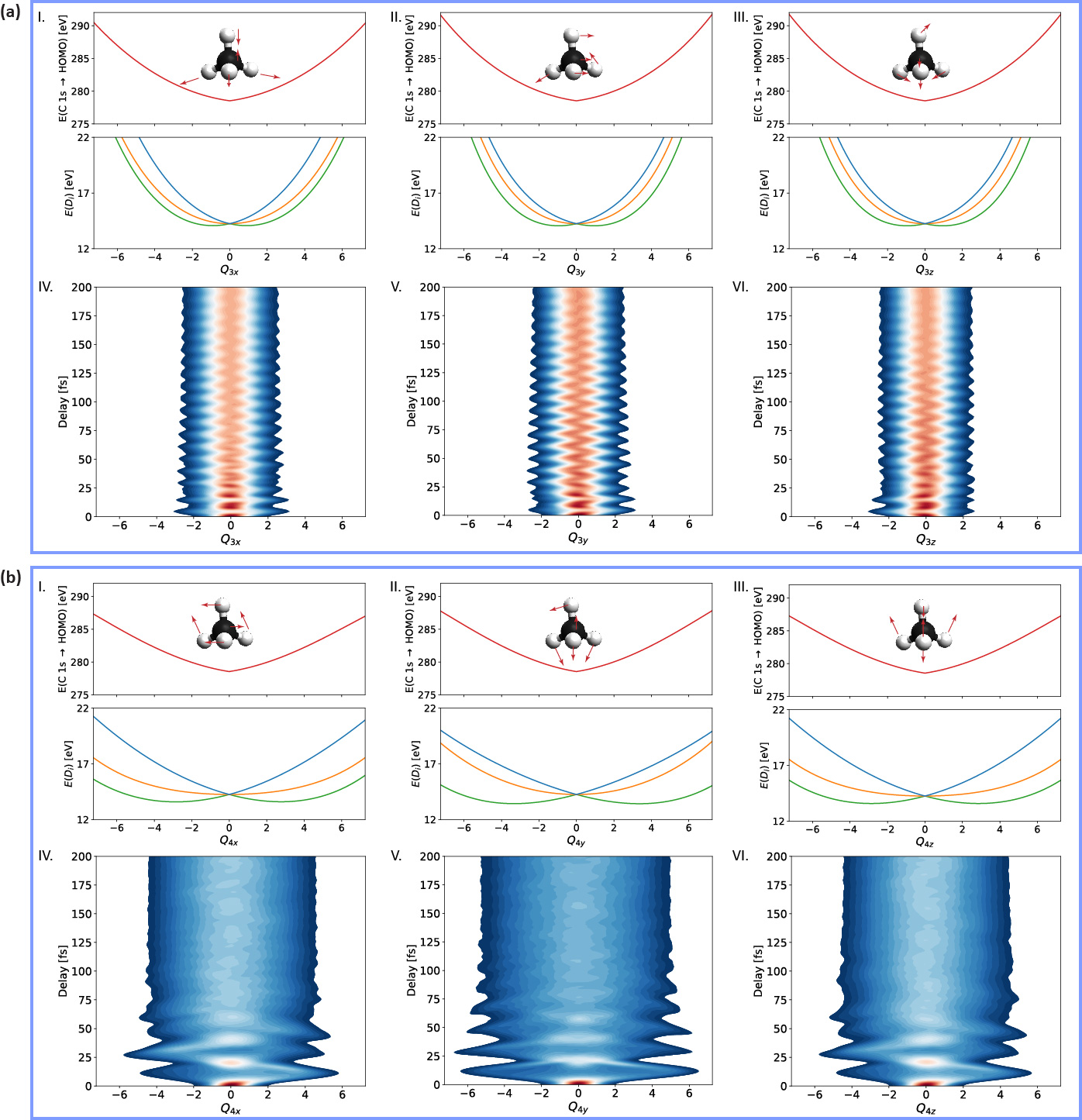}
    \caption{\textbf{Time-evolution of the nuclear wave packets along the other normal-mode coordinates}
    Same as Fig.~\ref{fig:3} for the $Q_3$ and $Q_4$ vibrational modes.
    \label{fig:3b}
    }
\end{figure}

We now discuss how these results allow us to understand the structural rearrangement CH$_4^+$ proceeding on a few-femtosecond time scale. The dominant dynamics observed in Fig.~\ref{fig:2} originate from the $Q_1$ symmetric stretching and the $Q_2$ scissoring mode. This conclusion follows from an analysis of the potential-energy surfaces and one-mode reduced nuclear densities calculated along the corresponding normal-mode coordinates displayed in Figs.~\ref{fig:3} and \ref{fig:3b}. The PES of CH$_4^+$ (central row) is displaced to positive values of the $Q_1$ symmetric-stretch coordinate which explains the creation of pronounced wave packet oscillations along this dimension. 
Since the energy of the core-excited PES increases along $Q_1$ (top row of Fig.~\ref{fig:3}) and the oscillator strength of the corresponding transition also varies with $Q_1$ (Fig.~\ref{fig:3c}), the 3100$\pm$153~cm$^{-1}$ frequency in the experimental data can be assigned to the symmetric stretch vibration $Q_1$. Along the $Q_{2x}$ coordinate, the PES of CH$_4^+$ splits into three components as a consequence of the JTE. Since the $Q_{2y}$ coordinate conserves $D_{\rm 2d}$ symmetry, the PES of CH$_4^+$ only splits into two components in this dimension. The large stabilization energies along both displacement coordinates cause large-amplitude wave-packet dynamics in both dimensions. Over the first ($\sim$20~fs) period of the $Q_2$ vibration, the nuclear wave packet stays relatively localized, while exploring the local minima of this PES along this e-symmetry vibration. Since the core-excited PES (top row of Fig.~\ref{fig:3}) increases by $\sim$7~eV over the excursion range of the $Q_2$ scissoring vibration and its $\sim$20~fs period matches the timescale observed in the spectral data (Fig.~2), it is clear that the scissoring dynamics dominate the early structural rearrangement of CH$_4^+$. 

\begin{figure}[!ht]
    \centering
    \includegraphics[width=0.7\textwidth]{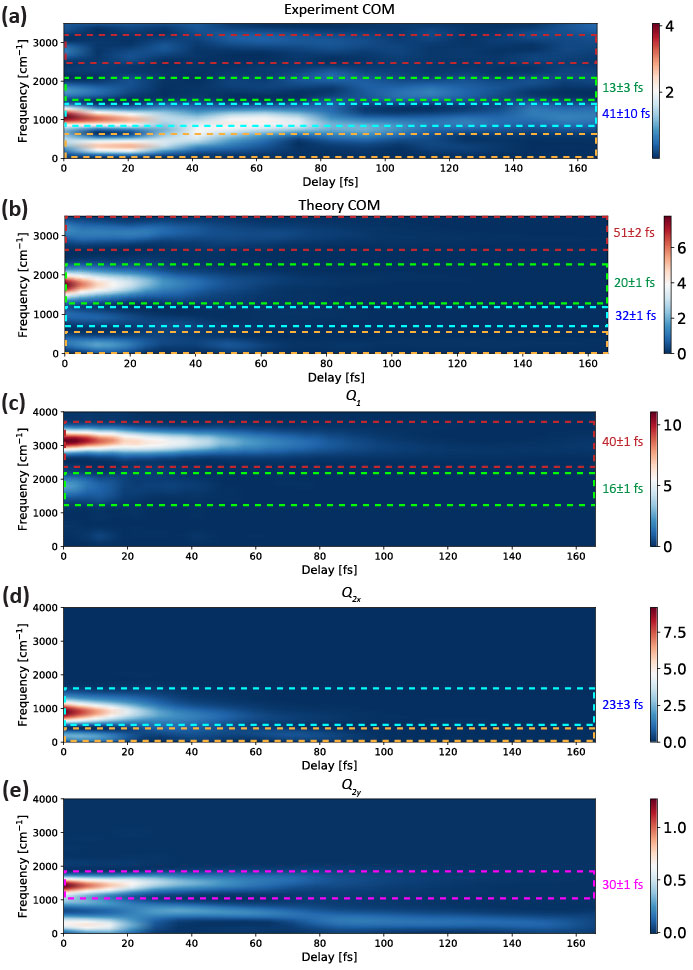}
    \caption{
    \textbf{Gabor transform spectrograms} of the center of mass of (\textbf{a}) the experimental and (\textbf{b}) the theoretical $\Delta$OD. (\textbf{c,d,e}) Spectrograms of the calculated center of mass of the time-dependent nuclear density along $Q_{1}$, $Q_{2x}$, $Q_{2y}$, respectively, shown in Fig. \ref{fig:3}.
    \label{fig:4}
    }
\end{figure}

Comparing the nuclear wave packet densities in Figs.~\ref{fig:3} and \ref{fig:3b}, it becomes apparent that the degree of wave-packet dispersion along different vibrational modes is very different, which has important implications for the structural-rearrangement dynamics of CH$_4^+$. To obtain additional insights and compare experiment and theory directly, we performed Gabor transformations of the center of mass (COM) of the C1s$\rightarrow$HOMO absorption band (details are given in Section \ref{experiment}), which are shown in Fig.~\ref{fig:4}\textbf{a} (experiment) and in Fig.~\ref{fig:4}\textbf{b} (theory). These spectrograms can be divided into four spectral regions highlighted by dashed boxes. The red and green boxes correspond to the frequency range of the symmetric-stretching vibration ($Q_1$), whereas the cyan and orange boxes correspond to frequencies of the scissoring vibrations ($Q_2$). All observed frequencies are damped, but on notably different time scales. Single-exponential fits, shown in Fig.~\ref{fig:4c}, yield decay constants of 13$\pm$3~fs (20$\pm$1~fs) for the stretching vibration in the green box for the experiment (theory), and 41$\pm$10~fs (32$\pm$1~fs) for the scissoring vibration in the cyan box for the experiment (theory).
Some of the frequency components displays oscillations, best visible in Fig.~\ref{fig:4c}, on top of an exponentially decaying component. 

\begin{figure}[t]
    \centering
    \includegraphics[width=0.8\textwidth]{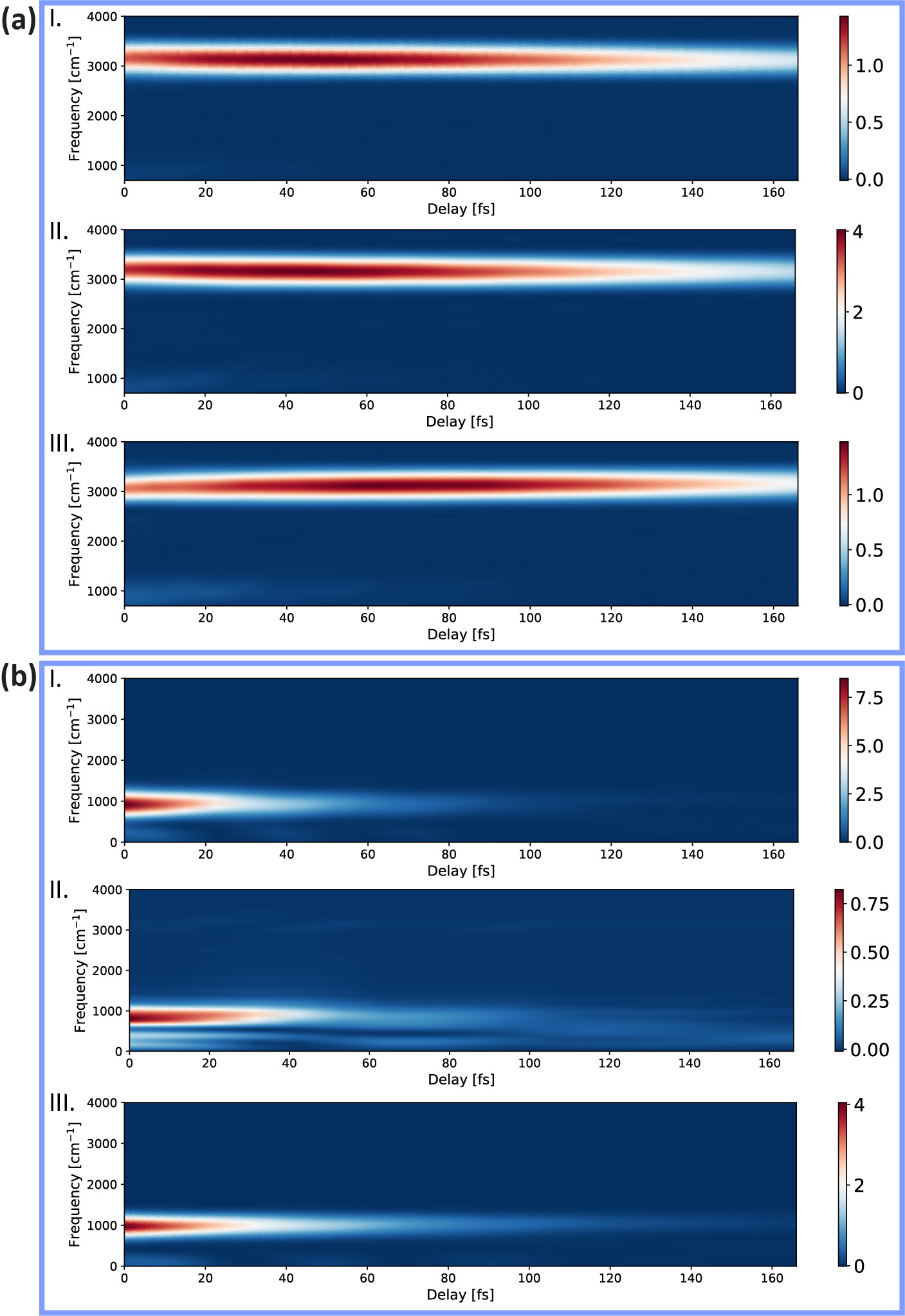}
    \caption{\textbf{Gabor-transform spectrograms}
     of the calculated center of mass of the time-dependent nuclear density along (\textbf{a}) the 3 coordinates (x,y,z) of the $Q_{3}$ asymmetric-stretch vibration and (\textbf{b}) the 3 coordinates (x,y,z) of the $Q_4$ scissoring vibration.
    \label{fig:4b}
    }
\end{figure}

To relate the damping of these vibrational frequencies to the structural rearrangement, we performed a similar analysis on the nuclear wave packet densities along each of the normal-mode coordinates (shown in Fig.~\ref{fig:4}\textbf{c-e}). This analysis is motivated by the fact that the oscillator strength of the C1s$\rightarrow$HOMO transition depends almost linearly on each of the normal-mode coordinates, as shown in Fig.~\ref{fig:3c}. Such a linear dependence suggests that modulations in the observable X-ray-absorption data can indeed be directly related to the underlying structural dynamics.
The comparison of the Gabor transforms of the center of mass of the nuclear wave packets (calculated from the data given in Fig.~\ref{fig:3}\textbf{d-f}) and the center of mass of the X-ray spectral data (Fig.~\ref{fig:4}\textbf{a} and \textbf{b}) further supports the notion that the observed dynamics are dominated by the $Q_1$ symmetric-stretch and the $Q_2$ scissoring modes. The decay of the symmetric-stretch frequency in the wave packet density (green box in Fig.~\ref{fig:4}\textbf{c}) indeed agrees well with the corresponding decay in Fig.~\ref{fig:4}\textbf{b}. 
The wave-packet dynamics along the $Q_{2x}$ coordinate gives rise to a $\sim$900-1000~cm$^{-1}$ frequency component also visible in the experimental and theoretical COM data (blue boxes). Finally, the wave-packet dynamics along the $Q_{2y}$ coordinate gives rise to a $\sim$1500~cm$^{-1}$ frequency component, which is less pronounced in the X-ray absorption data, but also to a $\sim$300~cm$^{-1}$ component, which is quite clearly visible in both the experimental and theoretical COM (orange boxes).

Overall, we thus find that the COM of both experimental and theoretical X-ray absorption data (Fig.~\ref{fig:4}\textbf{a,b}) show frequency components that decay within tens of femtoseconds and that all of these features can be well accounted for in terms of the $Q_1$ symmetric stretch and $Q_2$ scissoring vibrations, both in terms of the observed frequencies and of their decay dynamics.
Although the other vibrational modes are also excited through the ionization process, their Gabor transforms (shown in Fig.~\ref{fig:4b}) differ notably from those of the experimental and theoretical spectral data. The asymmetric-stretch vibration ($Q_3$) is indeed not significantly damped over the first 160~fs and the frequency components in the orange box of Fig.~\ref{fig:4} ($<$1000~cm$^{-1}$) are assigned to the $Q_2$ scissoring mode (Fig.~\ref{fig:3}\textbf{d} and \textbf{e}) because they only appear in the $Q_{4y}$ mode, where they are very weak (Fig.~\ref{fig:4b}).

\begin{figure}[t]
    \centering
    \includegraphics[width=\textwidth]{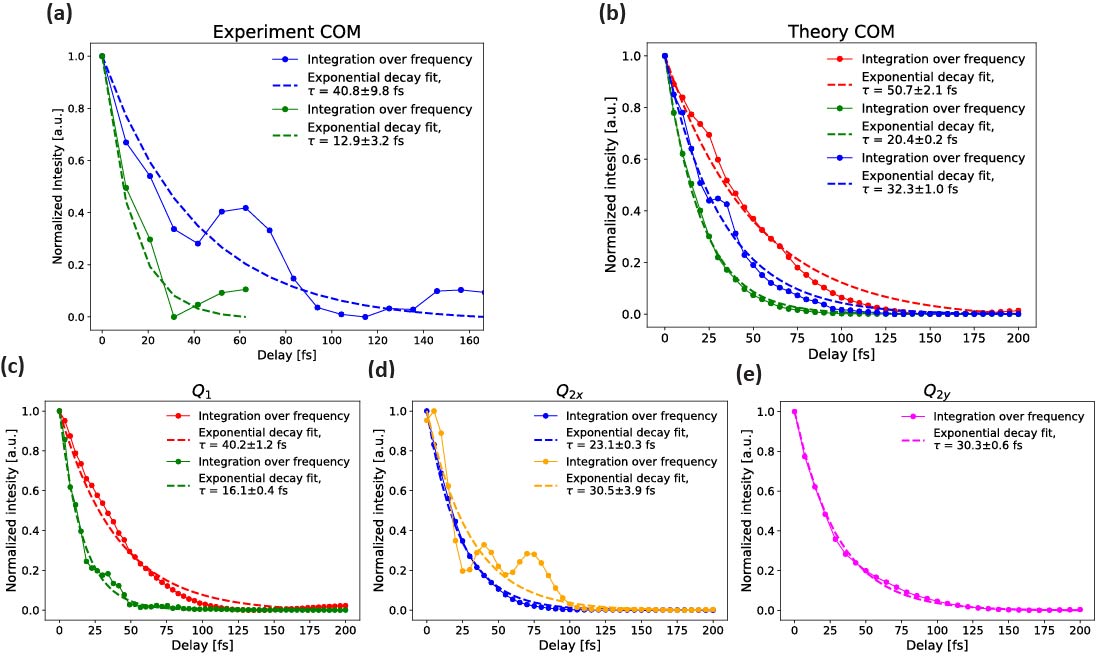}
    \caption{\textbf{Exponential fits of the vibrational frequency components in the Gabor transforms} of (a) the center of mass of the experimental $\Delta$OD signal from Fig.~2a, (b) the center of mass of the calculated $\Delta$OD signal from Fig.~2b, (c, d, e) the nuclear wave packet densities along the $Q_1$, $Q_{\rm 2x}$ and $Q_{\rm 2y}$ modes, respectively. The color of the data encodes the frequency region according to the dashed boxes shown in Fig.~\ref{fig:3}.
    \label{fig:4c}
    }
\end{figure}

\section{Discussion}

\begin{figure}[t]
    \centering
    \includegraphics[width=1\textwidth]{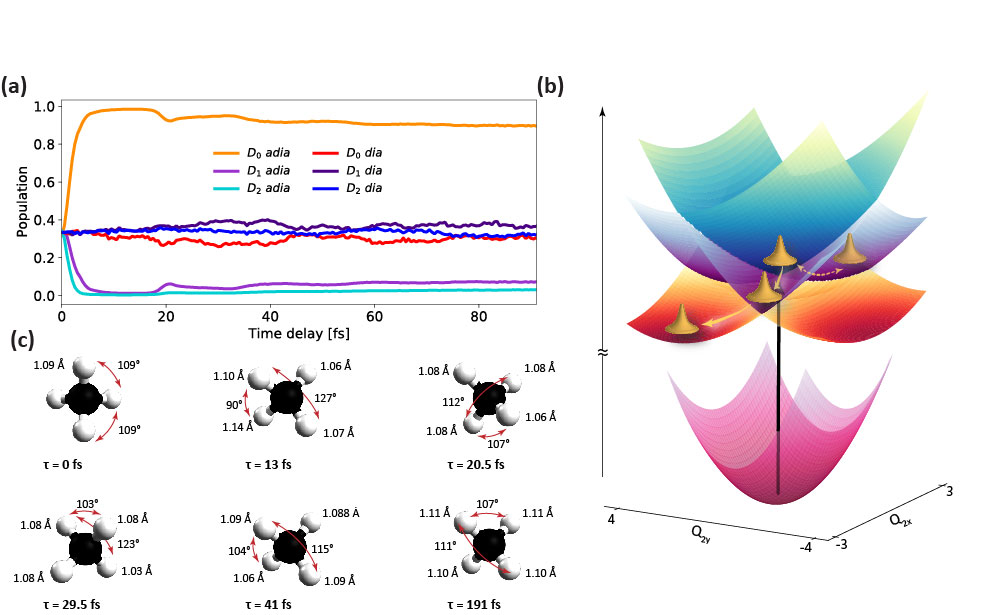}
    \caption{
    \textbf{Few-femtosecond electronic and structural rearrangement of CH$_4^+$.} \textbf{(a)} Population dynamics of the three lowest adiabatic electronic states of CH$_4^+$ and their diabatic counter-parts, as obtained from the 9-dimensional MCTDH calculations. \textbf{(b)} PES of the lowest three adiabatic states of CH$_4^+$ as a function of the two coordinates of the $Q_2$ scissoring vibration. \textbf{(c)} Snapshots of the average structure of CH$_4^+$ at the 6-time delays, whereby the four intermediate time delays were selected from the turning points of the spectral density shown in Fig.~\ref{fig:2}b. The largest and smallest bond angles are indicated in each case.
    \label{fig:5}
    }
\end{figure}

These experimental and theoretical results allow us to draw a picture of unprecedented detail of the electronic and structural rearrangements of CH$_4^+$. In our experiments, ionization of CH$_4$ by a few-cycle CEP-stable MIR pulse turns on the JTE in the ionized molecules within a fraction of a femtosecond close to the electric-field maximum of the pulse. This ionization step creates a wave packet centered on a seam of three-fold conical intersections that extends along the $Q_1$ symmetric-stretching mode. Along all other vibrational coordinates, the electronic degeneracy is lifted, as shown in Fig.~\ref{fig:3} and \ref{fig:3b}. Since CH$_4$ is randomly oriented in our experiments, all three initially degenerate components of the electronic ground state of CH$_4^+$ are equally populated at the time of ionization. Figure~\ref{fig:5}\textbf{a} shows that the populations of the higher-lying $D_1$ and $D_2$ adiabatic states relax into the D$_0$ adiabatic ground state in only 3.9$\pm$0.4~fs (from a mono-exponential fit of the early population dynamics) and that the higher-lying states are only repopulated to a few percent over the first 100~fs. Our interpretation of the dynamics of CH$_4^+$ therefore focuses on the structural rearrangement on the lowest cationic adiabatic state ($D_0$).

The analysis of the experimental and theoretical data has identified the $Q_1$ symmetric stretch and $Q_2$ scissoring modes as dominating the structural dynamics of CH$_4^+$. The comparison of the one-mode reduced nuclear densities (Fig.~\ref{fig:3} and \ref{fig:3c}) moreover reveals that a true structural relaxation only takes place along the $Q_{2y}$ vibrational coordinate, where the center of the wave packet density shifts from 0 to 2 in the dimensionless normal-mode coordinates. With the exception of a small shift to positive $Q_1$ in the long-time limit, we find no significant shifts of the center of the vibrational wave packet along any of the other normal-mode coordinates. This provides a clear picture of the wave-packet dynamics driving the structural rearrangement of CH$_4^+$, which is illustrated in the {$Q_{2x},Q_{2y}$} subspace in Fig.~\ref{fig:5}\textbf{b}. In this subspace, the nuclear wave packet starts at the position of the three-fold conical intersection at the instant of ionization. It undergoes oscillatory motion along the $Q_{2x}$ scissoring coordinate, which dephases in 23$\pm$3~fs, i.e. one period of the $Q_2$ vibration and leads to a symmetric spreading of the wave packet along this coordinate. This is indicated by the double-headed arrow along the $Q_{2x}$ dimension in Fig.~\ref{fig:5}\textbf{c}. 
The only true relaxation dynamics of CH$_4^+$ therefore take place along the $Q_{2y}$ dimension, which is unique because it causes a splitting of the PES that is asymmetric with respect to $Q_{2y}=0$. How this relaxation dynamics proceeds is illustrated by the full single-headed arrows in Fig.~\ref{fig:5}\textbf{c}.

These results allow us to construct a clear picture of the temporal evolution of the "structure" of CH$_4^+$ in the sense of the expectation values of the normal-mode coordinates. The corresponding geometries at time delays corresponding to the local extrema of the spectral positions as a function of time (as defined by the spectral data Fig.~\ref{fig:2}\textbf{a,b}) are shown in Fig.~\ref{fig:5}\textbf{c}. We find that the geometry of CH$_4^+$ corresponding to the COM of the wave packet is tetrahedrally symmetric at the instant of ionization ($t=0$) after which the dominant initial dynamics take place along the $Q_2$ scissoring mode that reduces one bond angle to 90$^{\circ}$ while increasing the other one to 127$^{\circ}$ by $\tau=13$~fs. We note that this geometry is still quite different from the $C_{\rm 2v}$ equilibrium geometry of CH$_4^+$ illustrated in Fig.~\ref{fig:1}, where the smallest bond angle amounts to 53$^{\circ}$ and the longest bond length is 1.17~\AA.
By $\tau=20.5$~fs both the bond angles and the bond lengths have returned close to their initial values because this delay corresponds to approximately one vibrational period of the scissoring modes and about two periods of the stretching modes. At $\tau=29.5$~fs, we again observe a notable difference between the largest and the smallest bond angles (104$^{\circ}$ vs. 123$^{\circ}$), as expected from the delay that now corresponds to $\sim$1.5 periods of the scissoring vibrations, but there is little difference in the bond lengths. For delays longer than 50~fs, we find little changes in either the spectral density (Fig.~2\textbf{a} and \textbf{b}) or the nuclear wave-packet densities (Fig.~3 and \ref{fig:3c}), and correspondingly, we find a nearly time-independent average structure of CH$_4^+$, which is illustrated at an exemplary delay $\tau=$191~fs in Fig.~\ref{fig:5}\textbf{c}. Importantly, this geometry is only weakly distorted compared to the initial tetrahedral geometry at $\tau=$0~fs with the smallest bond angle amounting to 107$^{\circ}$ and the largest one amounting to 111$^{\circ}$. This structure has $D_{\rm 2d}$ symmetry, which is consistent with our observation that the only true structural relaxation of CH$_4^+$ takes place along the $Q_{\rm 2y}$ coordinate, which preserves $D_{\rm 2d}$ symmetry. We thus conclude that CH$_4^+$ prepared by ionization of CH$_4$ electronically relaxes to the lowest of its adiabatic PES in 3.9$\pm$0.4~fs, where it remains structurally highly fluxional, but rearranges its average structure to a weakly distorted $D_{\rm 2d}$ geometry within 50~fs.

We note that these results are at odds with previous quantum dynamics simulations performed using MCTDH and a similar model Hamiltonian\cite{mondal15a}, in which it was predicted that a $C_{\rm 2v}$ structure was reached by the evolving wave packet. This disagreement, however, may be understood by the fact that the model potential of Reference~\citenum{mondal15a} contains no coupling terms between the totally symmetric stretch, $Q_1$, and the remaining $e$ and $t$ modes. These terms are not necessarily zero by symmetry, and their inclusion is important in order to correctly describe the structural dynamics. In the present model, this coupling is described to second-order via the bi-linear coupling coefficients $\eta_{1 \beta}^{(I,J)}$, $\beta \ne 1$, in Equation~\ref{eq:modpot}. These terms act to damp both the symmetric C-H stretch as well as the modes to which it couples. Through this mechanism, the adoption of a $C_{\rm 2v}$ structure in the long-time limit is inhibited.

Before concluding, we briefly compare and contrast our findings with the previous literature. On the basis of SFI pump, SFI probe experiments with 25-fs pulses and quantum dynamics from a two-dimensional model of the PES of CH$_4^+$, Lin et al. concluded that CH$_4^+$ prepared by ionization of CH$_4$ reached its $C_{\rm 2v}$ equilibrium geometry in 20$\pm$7~fs.
In Ridente et al., which employed experimental data similar to that presented here, the authors did not observe the stretching vibrational frequencies ($\sim$3100~cm$^{-1}$) detected in our work. Additionally, on the the basis of the comparison to classical-trajectory calculations, the authors concluded that CH$_4^+$ reached its C$_{\rm 2v}$ equilibrium geometry in 10$\pm$2~fs. They moreover concluded that the vibrational coherence of the initial scissoring motion was lost through internal vibrational redistribution into lower frequency modes in 58$\pm$13~fs. Our results show that CH$_4^+$ prepared by ionization of CH$_4$ never adopts its nominal C$_{\rm 2v}$ equilibrium geometry, but instead asymptotically adopts a geometry that is best described as only weakly D$_{\rm 2d}$ distorted. Specifically after 10$\pm$2~fs, both stretching vibrations have completed one period, such that the distortion is dominated by the scissoring modes only. Our results further show that the damping of a vibrational frequency observed by X-ray absorption cannot be directly interpreted as a loss of vibrational coherence, nor specifically be assigned to IVR. Our results indeed show that nuclear wave-packet spreading occurs to a similar extent along the $Q_{2x},Q_{4x},Q_{4y}$ and $Q_{4z}$ coordinates, and similarly, but asymmetrically along the $Q_{2y}$ as a consequence of the anharmonicity of the PES along these coordinates, caused by the JTE. No evidence of IVR between the $Q_2$ and the only lower-frequency mode ($Q_4$) was obtained, neither in our quantum-dynamical calculations, nor in our experiments.

\section{Conclusions and Outlook}

Combining ATAS experiments at the carbon K-edge with full-dimensional quantum-dynamics simulations coupled to X-ray absorption calculations, we have elucidated the few-femtosecond electronic and structural relaxation dynamics of methane cation driven by the Jahn-Teller effect. Our results show that the electronic relaxation proceeds in only 3.9$\pm$0.4~fs, followed by very large amplitude vibrational dynamics on the lowest adiabatic sheet of the PES, characterized by wave packet spreading within a few tens of femtoseconds. The structural relaxation dynamics are dominated by the $Q_2$ scissoring and the $Q_1$ symmetric stretching vibration, which dephase in 41$\pm$10~fs and 13$\pm$3~fs as a consequence of wave-packet dispersion, in reasonable agreement with theory (32$\pm$1~fs and 20$\pm$0.2~fs), respectively. Our results further show that CH$_4^+$ remains a highly fluxional species that possesses a time-averaged $D_{\rm 2d}$ structure because significant structural relaxation is restricted to the $Q_{\rm 2y}$ mode. This work demonstrates the considerable potential of ATAS and quantum-dynamics simulations to fully understand the fastest coupled electronic and structural rearrangements that occur in molecules, which holds considerable promise for understanding Jahn-Teller driven dynamics in larger molecules, such as fullerenes\cite{gunnarsson95a,chancey97a,dunn2005jahn}, metal complexes\cite{streltsov20a} and perovskites\cite{varignon19a}, both isolated or in solution\cite{yin23a}, as well as cooperative effects underlying the dynamics of strongly correlated materials\cite{iwahara13a,huang21a}. 

\section*{Supplementary Material}
The supplementary material contains additional information on the calculations reported in this manuscript.

\begin{acknowledgments}
We thank A. Schneider, M. Kerellaj, and M. Seiler for their technical support, J.-P. Wolf for fruitful discussions. Funding: HJW gratefully acknowledges funding from ERC Consolidator Grant (Project No. 772797-ATTOLIQ), and the Swiss National Science Foundation through project 200021\_172946 and the NCCR-MUST.
\end{acknowledgments}


\section*{Author Declarations}
The authors declare no competing interests.

\bibliography{phd,attobib}

\end{document}



\title{Supplementary material for: Few-femtosecond electronic and structural rearrangements of CH$_4^+$ driven by the Jahn-Teller effect} 



\author{Kristina S. Zinchenko}
\affiliation{Laboratory of Physical Chemistry, ETH Z\"{u}rich, 8093 Z\"{u}rich, Switzerland}
\author{Fernando Ardana-Lamas}
\affiliation{Laboratory of Physical Chemistry, ETH Z\"{u}rich, 8093 Z\"{u}rich, Switzerland}
\author{Valentina Utrio Lanfaloni}
\affiliation{Laboratory of Physical Chemistry, ETH Z\"{u}rich, 8093 Z\"{u}rich, Switzerland}
\author{Nicholas Monahan}
\affiliation{Laboratory of Physical Chemistry, ETH Z\"{u}rich, 8093 Z\"{u}rich, Switzerland}
\author{Issaka Seidu}
\affiliation{National Research Council of Canada, Ottawa, ON, Canada}
\author{Michael S. Schuurman}
\affiliation{National Research Council of Canada, Ottawa, ON, Canada}
\author{Simon P. Neville}
\email{simon.neville@nrc-cnrc.gc.ca}
\affiliation{National Research Council of Canada, Ottawa, ON, Canada}
\author{Hans Jakob Wörner}
\email{hwoerner@ethz.ch}
\affiliation{Laboratory of Physical Chemistry, ETH Z\"{u}rich, 8093 Z\"{u}rich, Switzerland}


\date{\today}

\pacs{}

\maketitle 

\section{MCTDH calculation details}

\begin{table}[h!]
    \centering
    \caption{Computational details of the MCTDH calculations. $N_{i}, N_{j}$ are the number of primitive harmonic oscillator DVR primitive functions used to describe each combined mode. $n_{i}$ are the number of single-particle functions used for each electronic state, in the order $|\tilde{X}_{x}^{+} \rangle$, $|\tilde{X}_{y}^{+} \rangle$, $|\tilde{X}_{z}^{+} \rangle$, $| \tilde{\mathcal{C}}^{+} \rangle$.}
    \begin{tabular}{lll}
    \hline
    Combined mode & $N_{i}, N_{j}$ & $n_{1}, n_{2}, n_{3}, n_{4}$ \\
    \hline
    $Q_{4x},Q_{3x}$ & 45, 11 & 16, 16, 16, 16\\
    $Q_{4y},Q_{3y}$ & 45, 11 & 16, 16, 16, 16\\
    $Q_{4z},Q_{3z}$ & 45, 11 & 16, 16, 16, 16\\
    $Q_{2x},Q_{2y}$ & 45, 45 & 16, 16, 16, 16\\
    $Q_{1}$       & 25     & 6,  6, 6, \\
    \hline
    \end{tabular}
\end{table}

\section{Simulated photoelectron spectrum}

\begin{figure}[h!]
    \includegraphics[width=0.9\textwidth]{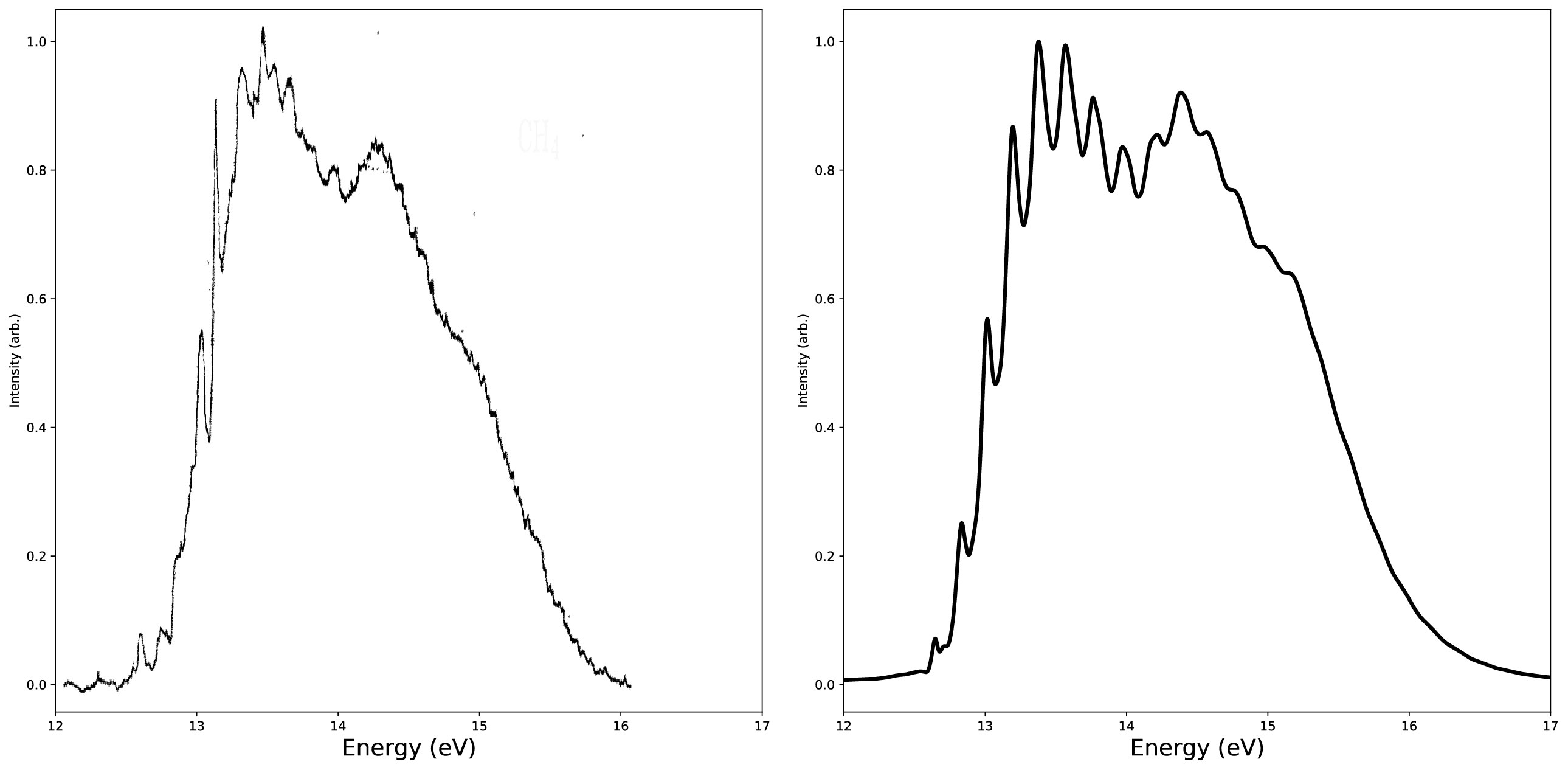}
    \caption{Left: experimetal photoelectron spectrum of Reference\citenum{potts1972}. Right: photoelectron spectrum computed via the Fourier transform of the wave packet autocorrelation function obtained following vertical displacement of the ground vibronic eigenstate to the $| \tilde{X}_{i}^{+} \rangle$ cation manifold and propagation using the model vibronic coupling Hamiltonian used to simulate the ATAS.}
    \label{fig:photoelc_spec}
\end{figure}

\section{Normal modes}
Given below are the ground state normal modes of CH$_{4}$ used in the MCTDH calculation, given in xyz format in units of $\AA$ as computed at the B3LYP/TZVP level of theory.

\begin{table}[h!]
    \centering
    \caption{$Q_{1}$}
    \begin{tabular}{ccccccc}
    \hline
    C &     0.0000000 &    0.0000000 &    0.0000000 &    0.0000000 &    0.0000000 &    0.0000008 \\
    H &     0.0000000 &   -0.8899253 &   -0.6292722 &    0.0000000 &   -0.4066333 &   -0.2875357 \\
    H &    -0.8899255 &    0.0000000 &    0.6292720 &   -0.4066375 &    0.0000000 &    0.2875409 \\
    H &     0.0000000 &    0.8899253 &   -0.6292722 &    0.0000000 &    0.4066333 &   -0.2875357 \\
    H &     0.8899255 &    0.0000000 &    0.6292720 &    0.4066375 &    0.0000000 &    0.2875409 \\
    \hline
    \end{tabular}
\end{table}

\begin{table}[h!]
    \centering
    \caption{$Q_{2x}$}
    \begin{tabular}{ccccccc}
    \hline
    C&      0.0000000&     0.0000000&     0.0000000&     0.0000000&     0.0000000&     0.0000000\\
    H&      0.0000000&    -0.8899253&    -0.6292722&    -0.4980267&     0.0000000&     0.0000000\\
    H&     -0.8899255&     0.0000000&     0.6292720&     0.0000000&    -0.4980267&     0.0000000\\
    H&      0.0000000&     0.8899253&    -0.6292722&     0.4980267&     0.0000000&     0.0000000\\
    H&      0.8899255&     0.0000000&     0.6292720&     0.0000000&     0.4980267&     0.0000000\\
    \hline
    \end{tabular}
\end{table}

\begin{table}[h!]
    \centering
    \caption{$Q_{2y}$}
    \begin{tabular}{ccccccc}
    \hline
    C&      0.0000000&     0.0000000&     0.0000000&     0.0000000&     0.0000000&     0.0000000\\
    H&      0.0000000&    -0.8899253&    -0.6292722&     0.0000000&     0.2875383&    -0.4066354\\
    H&     -0.8899255&     0.0000000&     0.6292720&     0.2875383&     0.0000000&     0.4066354\\
    H&      0.0000000&     0.8899253&    -0.6292722&     0.0000000&    -0.2875383&    -0.4066354\\
    H&      0.8899255&     0.0000000&     0.6292720&    -0.2875383&     0.0000000&     0.4066354\\
    \hline
    \end{tabular}
\end{table}

\begin{table}[h!]
    \centering
    \caption{$Q_{3x}$}
    \begin{tabular}{ccccccc}
    \hline
    C&      0.0000000&     0.0000000&     0.0000000&     0.0000000&     0.0000000&     0.0877021\\
    H&      0.0000000&    -0.8899253&    -0.6292722&     0.0000000&    -0.3960532&    -0.2612687\\
    H&     -0.8899255&     0.0000000&     0.6292720&     0.3960489&     0.0000000&    -0.2612657\\
    H&      0.0000000&     0.8899253&    -0.6292722&     0.0000000&     0.3960532&    -0.2612687\\
    H&      0.8899255&     0.0000000&     0.6292720&    -0.3960489&     0.0000000&    -0.2612657\\
    \hline
    \end{tabular}
\end{table}

\begin{table}[h!]
    \centering
    \caption{$Q_{3y}$}
    \begin{tabular}{ccccccc}
    \hline
    C&      0.0000000&     0.0000000&     0.0000000&     0.0000000&     0.0877030&    -0.0000000\\
    H&      0.0000000&    -0.8899253&    -0.6292722&     0.0000000&    -0.5413202&    -0.3960465\\
    H&     -0.8899255&     0.0000000&     0.6292720&    -0.0000000&     0.0187809&     0.0000000\\
    H&      0.0000000&     0.8899253&    -0.6292722&     0.0000000&    -0.5413202&     0.3960465\\
    H&      0.8899255&     0.0000000&     0.6292720&     0.0000000&     0.0187809&     0.0000000\\
    \hline
    \end{tabular}
\end{table}

\begin{table}[h!]
    \centering
    \caption{$Q_{3z}$}
    \begin{tabular}{ccccccc}
    \hline
    C&      0.0000000&     0.0000000&     0.0000000&     0.0877030&    -0.0000000&     0.0000000\\
    H&      0.0000000&    -0.8899253&    -0.6292722&     0.0187809&    -0.0000000&    -0.0000000\\
    H&     -0.8899255&     0.0000000&     0.6292720&    -0.5413202&    -0.0000000&     0.3960465\\
    H&      0.0000000&     0.8899253&    -0.6292722&     0.0187809&     0.0000000&    -0.0000000\\
    H&      0.8899255&     0.0000000&     0.6292720&    -0.5413202&    -0.0000000&    -0.3960465\\
    \hline
    \end{tabular}
\end{table}

\begin{table}[h!]
    \centering
    \caption{$Q_{4x}$}
    \begin{tabular}{ccccccc}
    \hline
    C &     0.0000000 &    0.0000000 &    0.0000000 &    0.0000000 &    0.1150320 &    0.0000000\\
    H &     0.0000000 &   -0.8899253 &   -0.6292722 &    0.0000000 &   -0.1291739 &    0.3019520\\
    H &    -0.8899255 &    0.0000000 &    0.6292720 &    0.0000000 &   -0.5562027 &    0.0000000\\
    H &     0.0000000 &    0.8899253 &   -0.6292722 &    0.0000000 &   -0.1291739 &   -0.3019520\\
    H &     0.8899255 &    0.0000000 &    0.6292720 &    0.0000000 &   -0.5562027 &    0.0000000\\
    \hline
    \end{tabular}
\end{table}

\begin{table}[h!]
    \centering
    \caption{$Q_{4y}$}
    \begin{tabular}{ccccccc}
    \hline
    C&      0.0000000&     0.0000000&     0.0000000&     0.0000000&     0.0000000&     0.1150314\\
    H&      0.0000000&    -0.8899253&    -0.6292722&     0.0000000&     0.3019506&    -0.3426913\\
    H&     -0.8899255&     0.0000000&     0.6292720&    -0.3019506&     0.0000000&    -0.3426913\\
    H&      0.0000000&     0.8899253&    -0.6292722&     0.0000000&    -0.3019506&    -0.3426913\\
    H&      0.8899255&     0.0000000&     0.6292720&     0.3019506&     0.0000000&    -0.3426913\\
    \hline
    \end{tabular}
\end{table}

\begin{table}[h!]
    \centering
    \caption{$Q_{4z}$}
    \begin{tabular}{ccccccc}
    \hline
    C&      0.0000000&     0.0000000&     0.0000000&    -0.1150320&     0.0000000&     0.0000000\\
    H&      0.0000000&    -0.8899253&    -0.6292722&     0.5562027&     0.0000000&    -0.0000000\\
    H&     -0.8899255&     0.0000000&     0.6292720&     0.1291739&     0.0000000&     0.3019520\\
    H&      0.0000000&     0.8899253&    -0.6292722&     0.5562027&    -0.0000000&    -0.0000000\\
    H&      0.8899255&     0.0000000&     0.6292720&     0.1291739&     0.0000000&    -0.3019520\\
    \hline
    \end{tabular}
\end{table}

\newpage
\clearpage
\bibliography{phd,attobib}